\documentclass{tcibook}
\usepackage{fancyhea}
\usepackage{work}
\usepackage{bm}       
\usepackage{graphicx}
\usepackage{multirow}
\usepackage{lineno}
\usepackage{threeparttable}
\usepackage[normalem]{ulem}
\usepackage{color}
\usepackage[colorlinks = true,
            linkcolor = blue,
            urlcolor  = blue,
            citecolor = blue,
            anchorcolor = blue]{hyperref}

\def\beq{\begin{equation}}
\def\eeq#1{\label{#1}\end{equation}}
\def\eeqn{\end{equation}}

\newenvironment{Eqnarray}%
   {\arraycolsep 0.14em\begin{eqnarray}}{\end{eqnarray}}
\def\beqa{\begin{Eqnarray}}
\def\eeqa#1{\label{#1}\end{Eqnarray}}
\def\eeqan{\end{Eqnarray}}

\let\bar=\overbar

\def\lsim{\mathrel{\raise.3ex\hbox{$<$\kern-.75em\lower1ex\hbox{$\sim$}}}}
\def\gsim{\mathrel{\raise.3ex\hbox{$>$\kern-.75em\lower1ex\hbox{$\sim$}}}}

\def\del{\partial}
\def\Dslash{\not{\hbox{\kern-4pt $D$}}}
\def\dslash{\not{\hbox{\kern-2pt $\del$}}}
\def\pslash{\not{\hbox{\kern-2pt $p$}}}
\def\ETmiss{\not{\hbox{\kern-4pt $E$}}_T}

\def\Dlr{\mathrel{\raise1.5ex\hbox{$\leftrightarrow$\kern-1em\lower1.5ex\hbox{$D$}}}}

\def\MSB{{\bar{M \kern -2pt S}}}
\def\msb{{\bar{\scriptsize M \kern -1pt S}}}

\def\drb{{\bar{\scriptsize D \kern -1pt R}}}

\def\authorlist#1#2{
    \vskip 0.4in
\begin{center}\begin{large} {\bf  #1 } \end{large}
    \vskip 0.2in
              #2
     \vskip 0.2in
   \end{center}
}

\begin{document}


\pagenumbering{roman}

\parindent=0pt
\parskip=8pt
\setlength{\evensidemargin}{0pt}
\setlength{\oddsidemargin}{0pt}
\setlength{\marginparsep}{0.0in}
\setlength{\marginparwidth}{0.0in}
\marginparpush=0pt



\renewcommand{\chapname}{chap:intro_}
\renewcommand{\chapterdir}{.}
\renewcommand{\arraystretch}{1.25}
\addtolength{\arraycolsep}{-3pt}











 \pagenumbering{arabic}





\chapter*{Snowmass Early Career}

\authorlist{Garvita Agarwal\textsuperscript{1}, Joshua L. Barrow\textsuperscript{2}, Mateus F. Carneiro\textsuperscript{3}, Thomas Y. Chen\textsuperscript{4}, Erin Conley\textsuperscript{5}, Rob Fine\textsuperscript{6}, Julia Gonski\textsuperscript{7}, Erin V. Hansen\textsuperscript{8}, Sam Hedges\textsuperscript{5}, Christian Herwig\textsuperscript{9}, Samuel Homiller\textsuperscript{10}, Tiffany R. Lewis\textsuperscript{11}, Tanaz A. Mohayai\textsuperscript{12}, Maria Elidaiana da Silva Pereira\textsuperscript{13}, Fernanda Psihas\textsuperscript{12}, Amber Roepe-Gier\textsuperscript{14}, Sara M. Simon\textsuperscript{9}, Jorge Torres\textsuperscript{15}, Jacob Zettlemoyer\textsuperscript{12}}
{1.University at Buffalo, State University of New York, Department of Physics, Buffalo, NY 14260, USA\\
2. Massachusetts Institute of Technology, Department of Physics, Cambridge, MA, USA 02139, USA\\
3. Brookhaven National Laboratory, Upton, NY 11973, USA\\
4. Fu Foundation School of Engineering and Applied Science, Columbia University, New York, NY 10027, USA\\
5. Duke University, Department of Physics, Durham, NC 27705, USA\\
6. Los Alamos National Laboratory, Los Alamos, NM, 87545, USA\\
7. Nevis Laboratories, Columbia University, Irvington, NY 10533, USA\\
8. University of California, Department of Physics, Berkeley, CA 94720, USA\\
9. Fermi National Accelerator Laboratory, Particle Physics Division, Batavia, IL 60510, USA\\
10. Department of Physics, Harvard University, Cambridge, MA, 02138, USA\\
11. NASA Postdoctoral Program Fellow, Goddard Space Flight Center, Greenbelt, MD 20771, USA\\
12. Fermi National Accelerator Laboratory, Neutrino Division, Batavia, IL 60510, USA\\
13. Hamburger Sternwarte, Universit{\"a}t Hamburg, Gojenbergsweg 112, 21029 Hamburg, Germany\\
14. University of California Santa Cruz, Physics Department, Santa Cruz, CA 95064, USA\\
15. Yale University, Wright Lab, New Haven, CT 06511, USA}

The Snowmass 2021 strategic planning process provided an essential opportunity for the United States high energy physics and astroparticle (HEPA) community to come together and discuss upcoming physics goals and experiments.
As this forward-looking perspective on the field often reaches far enough into the future to surpass the timescale of a single career, consideration of the next generation of physicists is crucial.

The 2021 Snowmass Early Career (SEC) organization aimed to unite this group, with the purpose of both educating the newest generation of physicists while informing the senior generation of their interests and opinions.
SEC is the latest in a series of the previously dubbed ``Snowmass Young" organizations, from 2013~\cite{Anderson:2013fca} and 2001~\cite{sm2001}.
This iteration has expanded on these efforts to significantly increase involvement and broaden the representation of the early career community in the process.

Early career physicists are the future of the field. 
They will design, build, and operate next-generation experiments and usher in new discoveries.
They are also disproportionately involved in work to develop an inclusive climate within HEPA.
This document summarizes the work of SEC in consolidating a huge variety of physics perspectives and community opinions towards a bright, strategic future.

\tableofcontents

\section{Executive Summary}
\label{sec:execsum}
Early career members of the HEPA community are a distinct population with a particular vision, perspective, and set of accompanying challenges that differ from those of the senior members of the field. Whether early career is defined as students and post-doctoral-equivalent stages or it includes faculty in their early career, what they have in common is a lower, often more vulnerable position in the hierarchy of our field. This, coupled with the fact that they are the source of a majority of the labor of our research, makes them placed to both uniquely contribute to and be affected by the direction of the field in the coming decade. 

The definition of early career used for this document is “people within 10 years of their most recent degree with time allowance for any long-term leave from the field”. In practice, an overwhelming majority of the participants in the SEC group and those identifying with the term Early Career are at a graduate or post-graduate (e.g. postdoc) equivalent career stage.

The early career community has made significant strides in broadening representation in HEPA since the last Snowmass process, particularly in their efforts to start and expand the reach of Early-Career Organizations (ECOs) within experimental collaborations and at national labs. These ECOs have played significant roles to identify the issues that are especially relevant to this community, and they have made substantial contributions to the betterment of our scientific reach as a community throughout the years.


The genesis of SEC for Snowmass 2021 came from a broadly distributed call for interest that returned over 250 responses. 
The organization was built to be fully inclusive, such that anyone who wanted to participate could do so. 
The definition of ``early career" (EC) that was adopted covered anyone with ten years of receiving their highest degree, with flexibility on this number to accommodate nontraditional career paths or breaks. 
This also was chosen to include technicians, engineers, and other members of the physics community that are not PhD holders.
Ultimately, a large majority of SEC participants were students and postdocs.
The current and previous Early Career members of the Division of Particles and Fields (DPF) Executive Committee were tasked with contributing to the leadership of this organization, ensuring close ties and clear communication with the organizing unit. 

\begin{figure}
    \centering
    \includegraphics[width=0.9\textwidth]{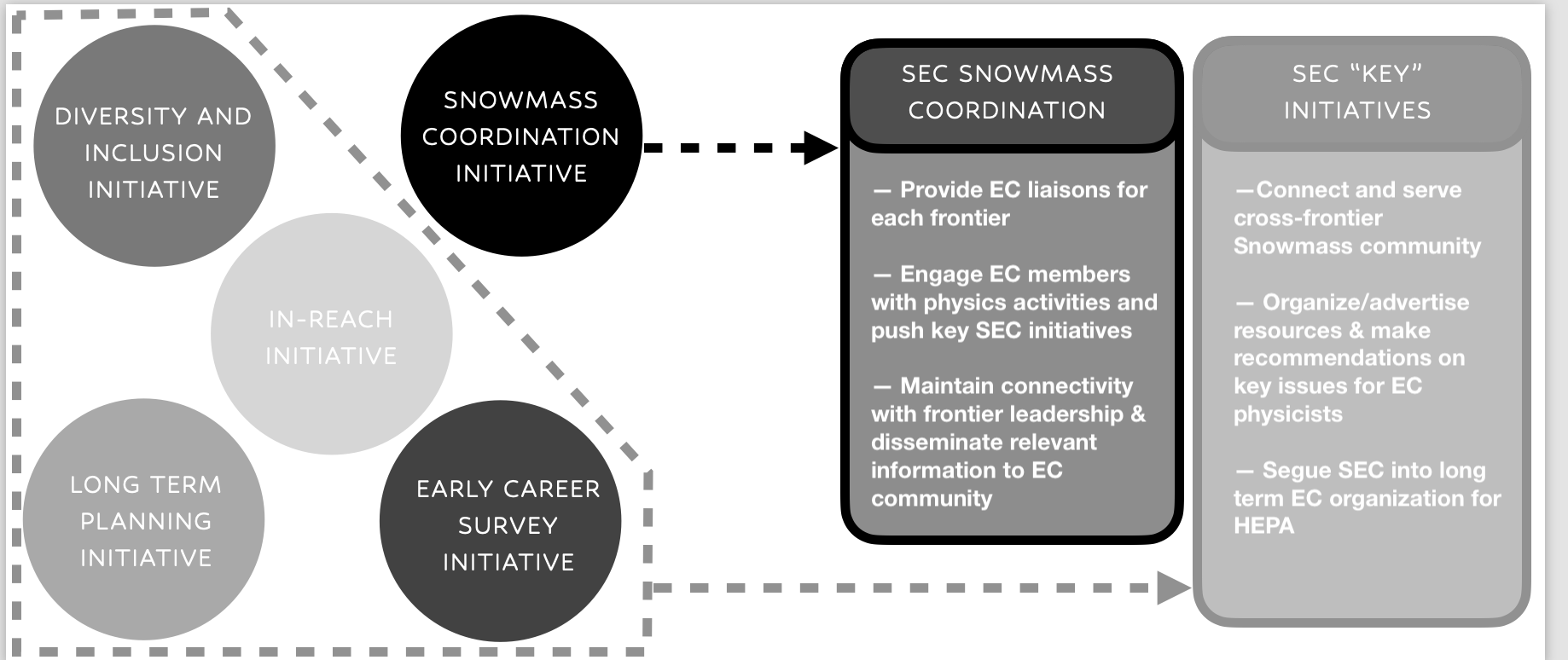}
    \caption{Organization of SEC. Left: The SEC sub-groups or initiatives, corresponding to spheres of early career interest. Right: Organizational division and development of ``Key" or ``Core" initiatives.}
    \label{fig:secstructure}
\end{figure}

Personpower then organically separated into two parallel groups, the structure of which is shown diagrammatically in Figure~\ref{fig:secstructure} and described in the text below.
The first was referred to as Snowmass Coordination, and consisted of liaisons to each frontier of the Snowmass structure.
Its goal was to connect early career (EC) physicists to projects within their frontiers of interest.
Liaisons had the responsibility of attending frontier-level meetings and representing the interests of the EC community.
Each frontier organized its liaison structure differently, depending on the number of candidates interested in a liaison position. The most common model was 1-3 liaisons with staggered rotating leadership to provide continuity while lessening the leadership commitment.

The second pillar of SEC was the Core Initiatives Organization, which gathered community input on the topics of greatest importance to EC scientists and set up subgroups dedicated to each. 
The four topics covered were Inreach, focusing on peer networking and connectivity among EC community members; Diversity, Equity \& Inclusion (DEI); Survey, to obtain up-to-date data on community opinions and perspectives; and Long-Term Organization, to discuss sustainability of the effort. 
The activities of these subgroups included the organization of informational panels and colloquia and the generation of recommendations for the general Snowmass management. 
The SEC-organized community-wide survey is one of the main P5-relevant works to come from the SEC organization, and includes information on careers, physics outlook, workplace culture, and US visa policies. The survey results are documented in detail in an associated white paper~\cite{survey}.
The key takeaways and recommendations from the SEC Survey are summarized in Section~\ref{sec:survey}.

More details on the creation and structure of the SEC organization can be found in a dedicated white paper~\cite{keyinit}.
This process elapsed over nearly two years due to delays from the COVID-19 pandemic, and the effect of this on efficacy and connectivity is discussed in the white paper.

\clearpage

\textbf{Through the Snowmass process, we developed a number of recommendations with input from the early career community.} Below we give a top-level summary of these recommendations. More detailed recommendations and their motivations are described in detail in the remainder of this chapter.

\begin{enumerate}
\item Institutions, experiments, and funding agencies should increase their commitment to adding and maintaining early career representation in decision-making bodies at all levels (e.g. review and advisory panels, governing bodies, etc.) and foster the development of early career organizations.
\item Institutions and funding agencies should address the need for economic equity for early career scientists by increasing the pay of early career positions to match industry equivalents and providing funds for improving meeting accessibility.
\item Funding agencies, experiments, and institutions should restructure the processes for reporting and investigation of discrimination and harassment to ensure true accountability and to robustly support equity, diversity, and inclusion in the field.
\item Career development and community efforts like outreach, mentoring, and advocacy should be recognized by institutions and funding agencies as critical tasks to the scientific output and health of the field. Institutions and funding agencies should provide support for service efforts, include this work in job expectations, ensure that faculty and scientists are given adequate time and credit for this work, and ensure that service work is equitably distributed.
\item Institutions should track career outcomes and adequately train early career scientists to move into a variety of job sectors, especially industry positions, through providing professional development opportunities, creating networking opportunities, and exposing early career scientists to a larger diversity of job sectors and mentors.
\item Institutions and funding agencies should continually examine and adapt their policies to address changing trends in HEPA, including changing job expectations, flexibility in remote work, and increasing competition in both the job market and funding opportunities.
\item Institutions and funding agencies should improve support for scientists with caregiving responsibilities, including encouraging reasonable work hours, providing adequate salaries, offering paid Medical and Family leave to all employees and supporting employees who use it, subsidizing or offering childcare, and fairly evaluating caregivers' drop in productivity in the context of current events of broad impact (e.g. COVID-19) in hiring and promotion committees.
\item Institutions and funding agencies should evaluate and assess the impacts of the COVID-19 pandemic over the coming years and adapt policy to support those most affected by COVID-19 and future events of similarly broad impact.
\item Institutions should provide comprehensive support to early career scientists, including resources, protections, and policies to support a healthy workplace culture and mental health
\item Institutions and funding agencies should take steps to improve U.S. visa and immigration policies by implementing more inclusive hiring processes as well as advocating for updated policies and streamlined application processes for scientists and STEM professionals.
\end{enumerate}

\clearpage

\section{SEC Survey Report}
\label{sec:survey}
The Snowmass Community Survey was designed by the Snowmass Early Career (SEC) Survey Core Initiative team between April 2020 and June 2021 with the aim of collecting demographic, career, physics outlook, and workplace culture data on a large segment of the Snowmass community. The team reviewed questions from past Snowmass surveys, developed new topics, and came to a consensus on the survey questions. The survey was released to the community on June 28, 2021 and had nearly $1500$ total interactions before it closed on August 26, 2021. This section summarizes the key findings and recommendations from the SEC survey. Most of the text in this section has been excerpted and paraphrased from the SEC survey report~\cite{survey} with permission from the authors.

The survey was distributed broadly via: the Snowmass Slack and email; an article in Fermilab At Work \cite{FNAL_SurveyAdvertisement}; an email to the Fermilab users list; collaboration spokespeople to their collaboration members; the July 2021 DPF newsletter \cite{APS_DPF_2021JulyNewsletter}; and advertisements on software forums (e.g. Geant4 and ROOT), American Physical Society (APS) forums, and social media sites (e.g. Reddit \cite{Reddit}). Qualtrics fraud detection features were used to detect duplicate submissions and screen for bots \cite{Qualtrics_Fraud}~\footnote{No submissions were tagged as duplicate. A few survey interactions were flagged as potential bots, but they were found to contain logical responses and comments and thus were not removed.}.

\begin{figure}[hbtp]
    \includegraphics[scale=0.56]{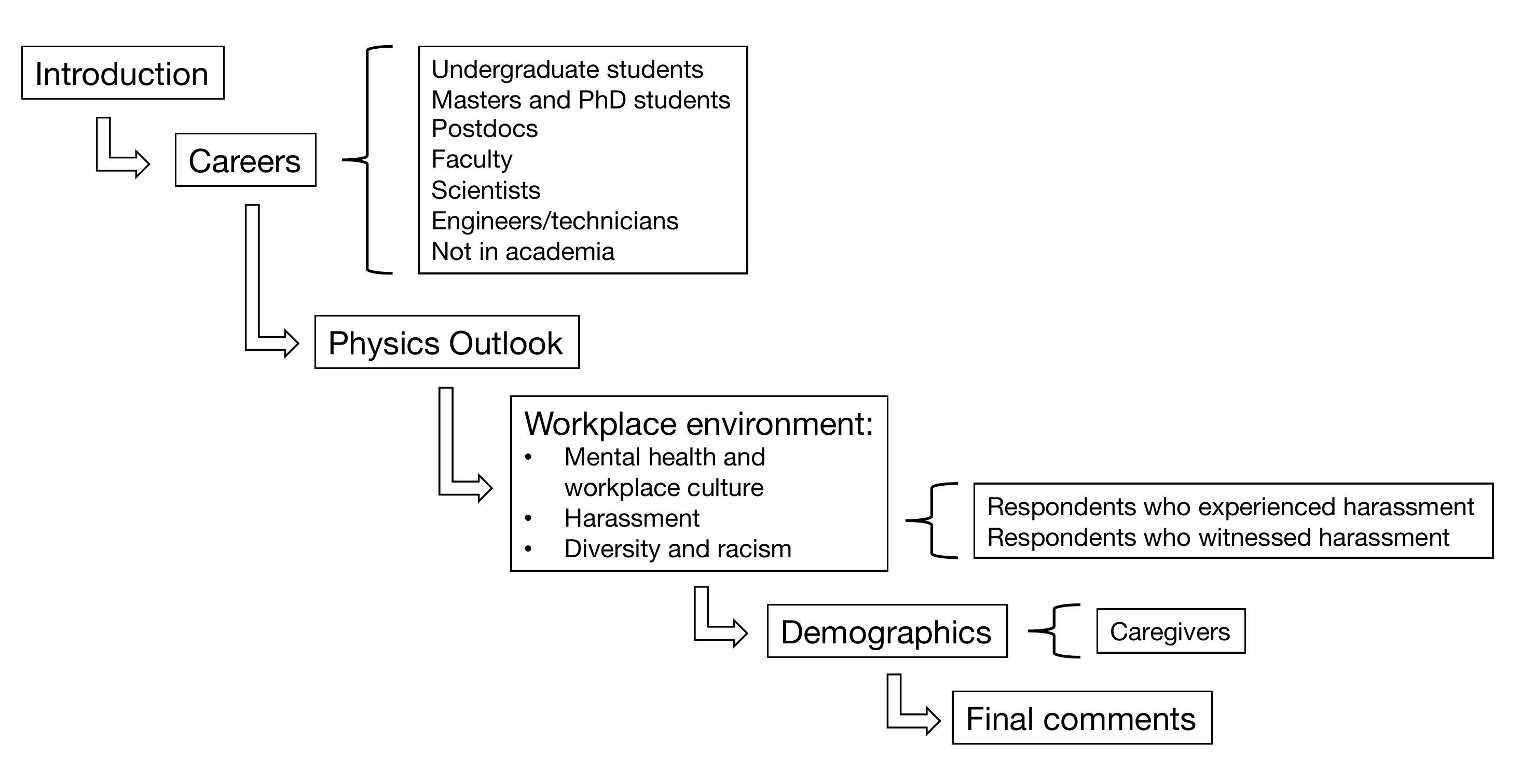}
    \caption{General overview of the 2021 Snowmass Community Survey. The arrows represent the overall flow of the survey ordered by the questions respondents viewed earliest, and the brackets present some of the display logic utilized in the survey.}
    \label{snowmass_survey_flowchart}
\end{figure}

The survey questions broadly fall into seven categories: demographics, physics outlook, careers, workplace culture, diversity and racism, caregiving responsibilities, and the impacts of COVID-19. Figure \ref{snowmass_survey_flowchart} shows the outline of the 2021 Snowmass Community Survey. The survey was implemented using online Qualtrics~\cite{Qualtrics} software hosted by Duke University, which enabled using display logic to ask unique sets of questions to different groups of respondents (represented by the brackets in Fig. \ref{snowmass_survey_flowchart}). Additionally, all questions except the first question (``Are you currently in academia?'') were optional, so questions did not have a set number of respondents. A majority of the physics outlook questions were only shown to respondents who answered that they were currently in academia.

Several measures were taken in the analysis to protect the respondents' privacy, including removing raw response numbers and re-normalizing distributions for all non-demographic results, only showing category breakdowns with $\geq 5$ responses in figures, and only separating responses based on provided demographic information into groupings with at least 30 responses. 

There were 1462 total interactions with the survey. The survey team limited the number of respondents to those that reached the end of (or continued beyond) the first section, which contained career-related questions. This reduced the total number of respondents to 1014.

Here we present a high level summary of the key findings and recommendations from the survey report. The full analysis and additional findings are available in the full survey report~\cite{survey}. Section~\ref{sec:physics_outlook} summarizes the key findings on physics outlook from the survey respondents, while Section~\ref{sec:survey_rec} discusses the key findings and recommendations of the other survey sections.

\subsection{Physics Outlook}
\label{sec:physics_outlook}
\begin{figure}[h!]
    \centering
    \includegraphics[width=1.0\linewidth]{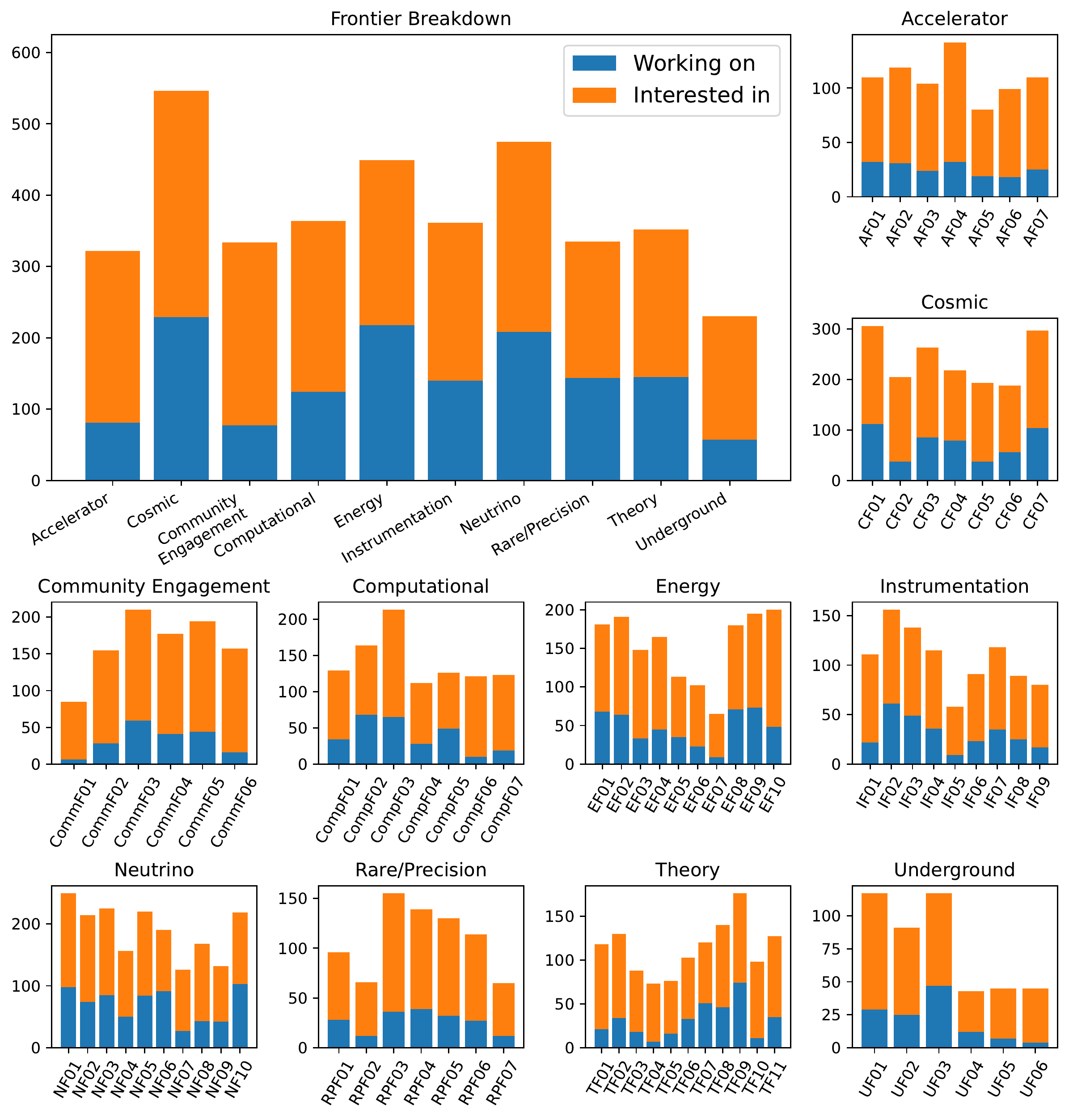}
    \caption{The survey team asked respondents to indicate which Frontiers and Topical Groups they were working in or had an interest in, even if they were not directly working on such physics.}
    \label{fig:frontier_breakdown}
\end{figure}

The survey responses demonstrate broad interest and participation across all of the different Frontiers, with a significant number of respondents indicating that they had interests outside their particular research focus \autoref{fig:frontier_breakdown}. Additionally over $40\%$ of respondents selected more than one Frontier as the Frontier they work in, indicating much of the research in HEPA extends beyond a single primary area.

The survey team made a deliberate effort to not ask for opinions on specific experiments. Because the field of high energy physics moves quickly, the choices given in previous surveys were not particularly relevant today. Additionally, there was concern that this type of question would simply reflect the relative number of participants from different experiments involved in Snowmass.

\begin{figure}[h!]
    \centering
    \includegraphics[width=\linewidth]{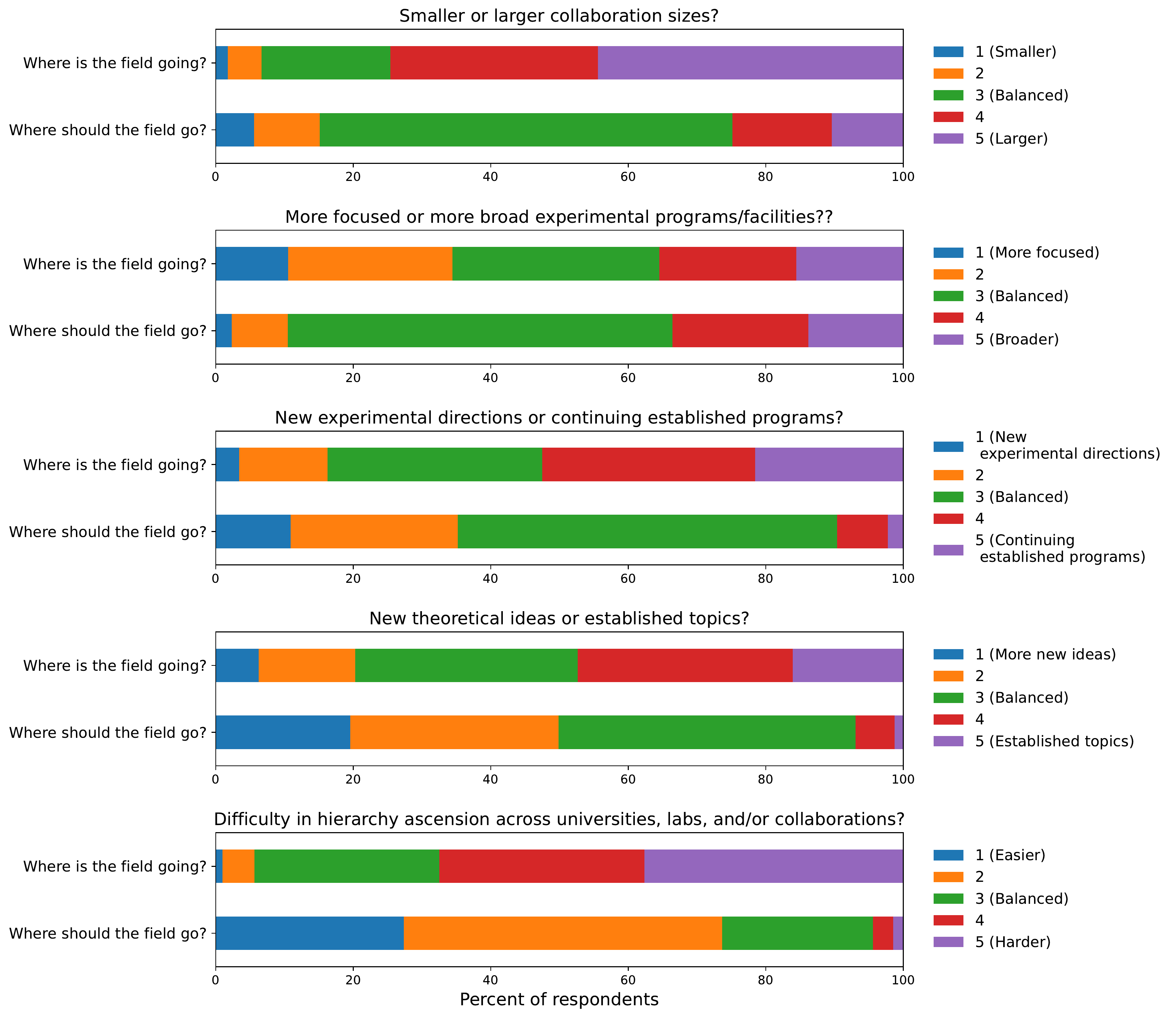}
    \caption{For a range of topics, respondents were asked to rate (on a 1-5 scale) in which direction they thought the field was currently going and where they thought the field should go.}
    \label{fig:physics_directions}
\end{figure}

Instead, respondents were asked more general questions about where they felt the field \textit{was moving} and in which direction they felt the field \textit{should be moving} \autoref{fig:physics_directions}. A plurality of respondents indicated they felt the field ``should be going'' for a balanced approach between small and large collaboration sizes, focused and broad experimental programs and facilities, new and continuing directions and programs, and new and established topics. Overwhelmingly, the biggest imbalance between where respondents felt the field was going compared to where it should be going was in the difficulty in hierarchy ascension. Nearly $70\%$ of respondents said the field was heading towards a 4 or 5 (with higher numbers implying more difficulty), while there was an overwhelming preference (with $> 70\%$) among respondents for this to become easier.

\begin{figure}[h!]
    \centering
    \hskip 2cm 
    \includegraphics[width=0.8\linewidth]{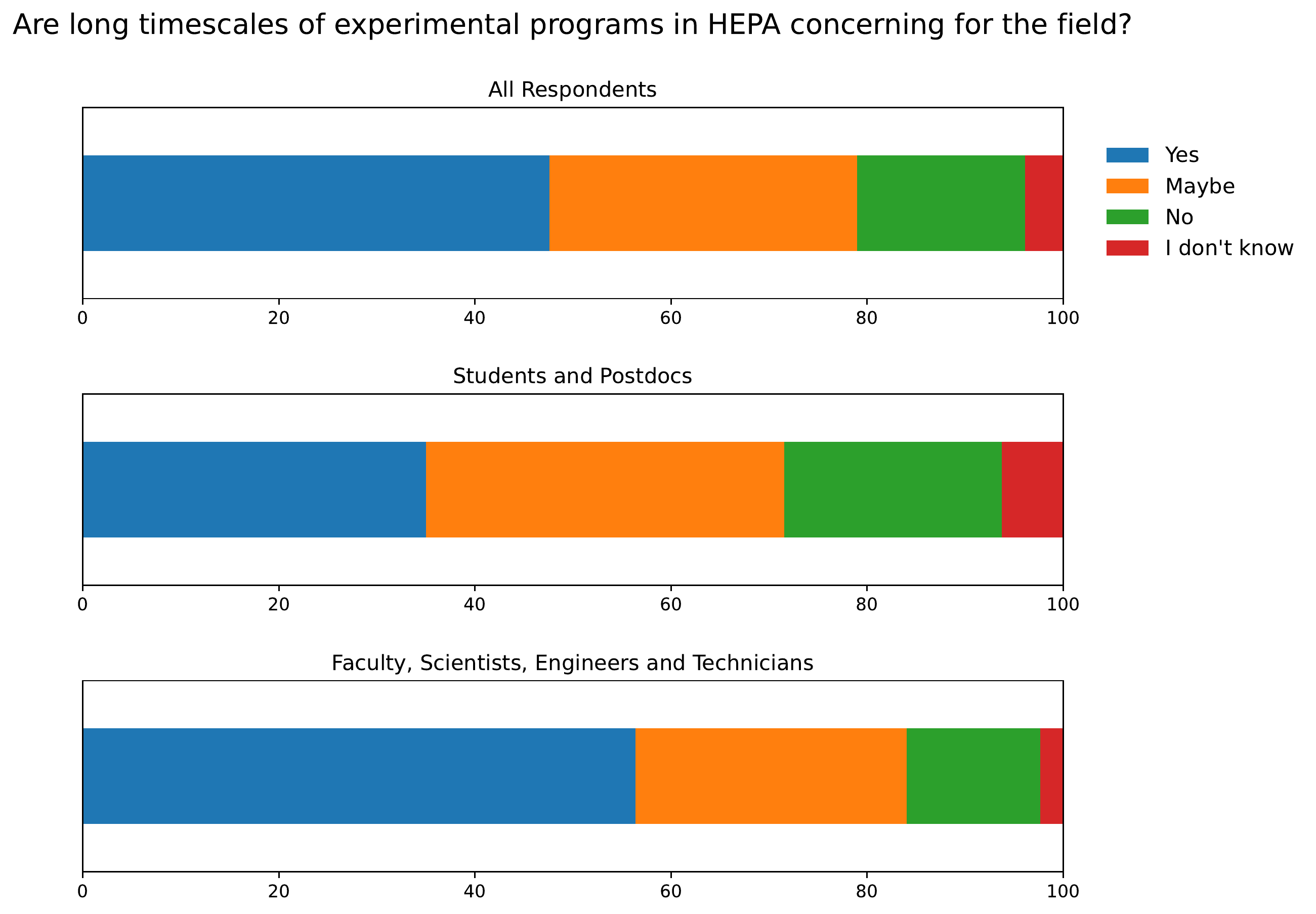}
    \caption{Respondents were asked whether they believed long timescales of experimental programs in HEPA were a concern for the field.}
    \label{fig:physics_timescales}
\end{figure}

Large-scale experimental projects in HEPA often span decades and even funding agencies, which can cause unique challenges to the careers of scientists in the field. When asked if they believed long timescales of experimental programs in HEPA were concerning for the field, a plurality of respondents answered ``Yes'', with another $\sim 30\%$ of respondents indicating ``Maybe''. Only about $20\%$ of respondents answered a definitive ``No'' \autoref{fig:physics_timescales}.

\begin{figure}[h!]
    \centering
    \includegraphics[width=0.9\linewidth]{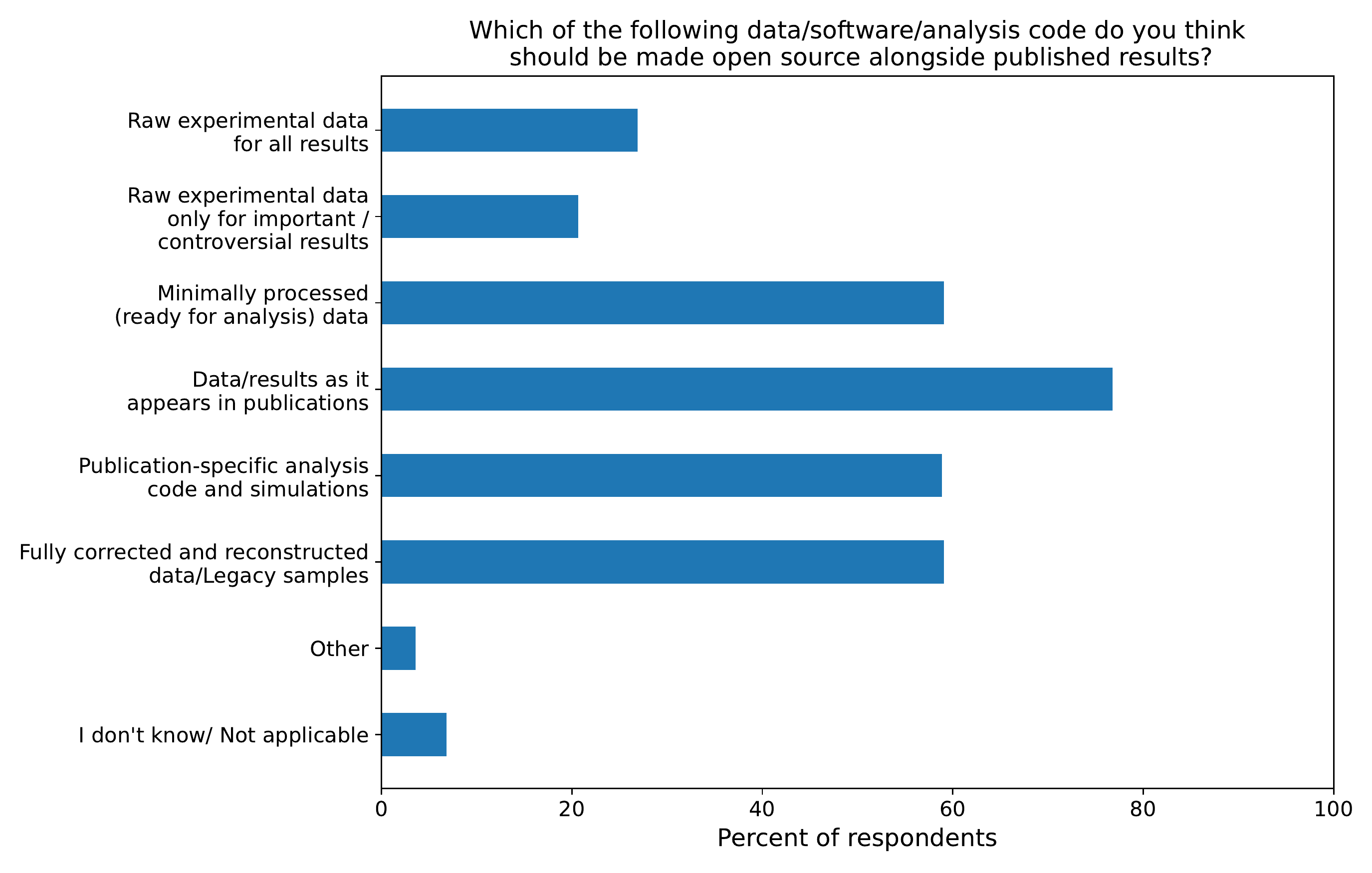}\hskip 0.5cm
    \caption{Respondents were asked which data/software/analysis code they believed should be made open source alongside published results.}
    \label{fig:outlook_6}
\end{figure}

The survey also asked respondents to select what types of data, software or analysis codes they believed should be made open source alongside published results. Most respondents were in favor of publishing data/results as they appear in publications, but a significant majority of respondents also said they supported ``Minimally processed (ready for analysis) data'',  ``Publication-specific analysis code and simulations'', and ``Fully corrected and reconstructed data / legacy samples'' being made open source \autoref{fig:outlook_6}.

\begin{figure}[h!]
    \centering
    \includegraphics[width=0.9\linewidth]{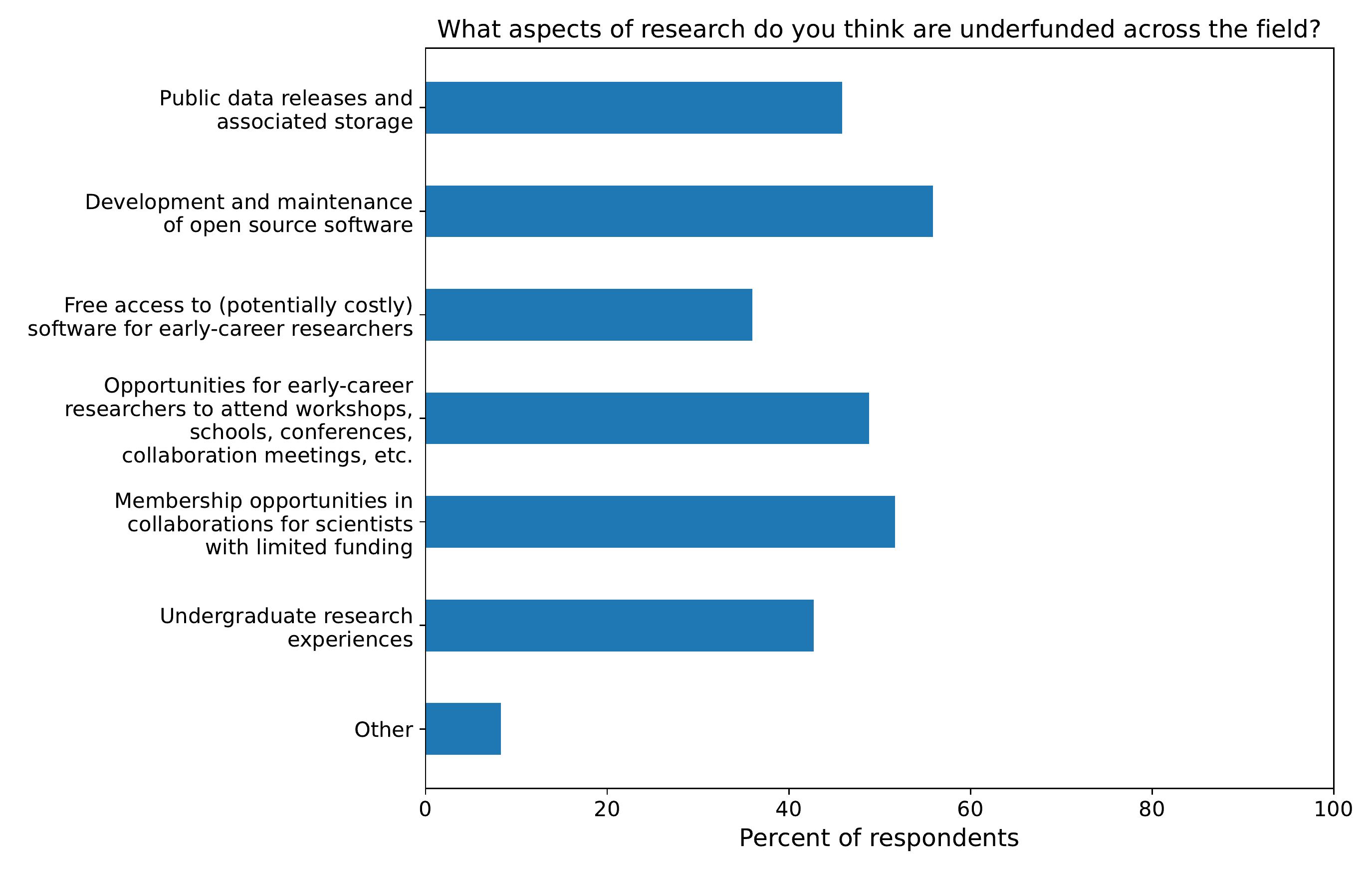}\hskip 0.5cm 
    \caption{Respondents were asked what aspects of research in HEPA they believed were underfunded.}
    \label{fig:physics_underfunded}
\end{figure}

When asked about which aspects of research in HEPA they believed were underfunded, the most frequently selected option was ``development and maintenance of open source software'', which over half of respondents selected.  Over $40\%$ of respondents also selected ``public data releases and associated storage'', ``Opportunities for early-career researchers to attend workshops, schools, conferences and meetings'', ''Membership opportunities in collaborations for scientists with limited funding'', and ``undergraduate research experiences'' \autoref{fig:physics_underfunded}. 

There has also been a growing awareness within the HEPA community about the climate emergency in society~\cite{climate}. When asked to what extent they know about the HEPA community’s environmental impact, on average, respondents said they do not know much about it. Early career scientists report knowing slightly less than senior researchers. The survey team also asked how concerned the respondents were about the HEPA’s environmental impact. The responses are, on average, more neutral. Nevertheless, early career members seem to be more concerned than senior researchers. 
Finally, the respondents think, on average, that it is important to take into account the environmental impact when making decisions on future HEPA projects. 
This particularly emerges from early career researchers, who will most likely cope with more severe effects of the climate emergency.

\begin{figure}[h!]
  \centering
  	\includegraphics[width=1.0\linewidth]{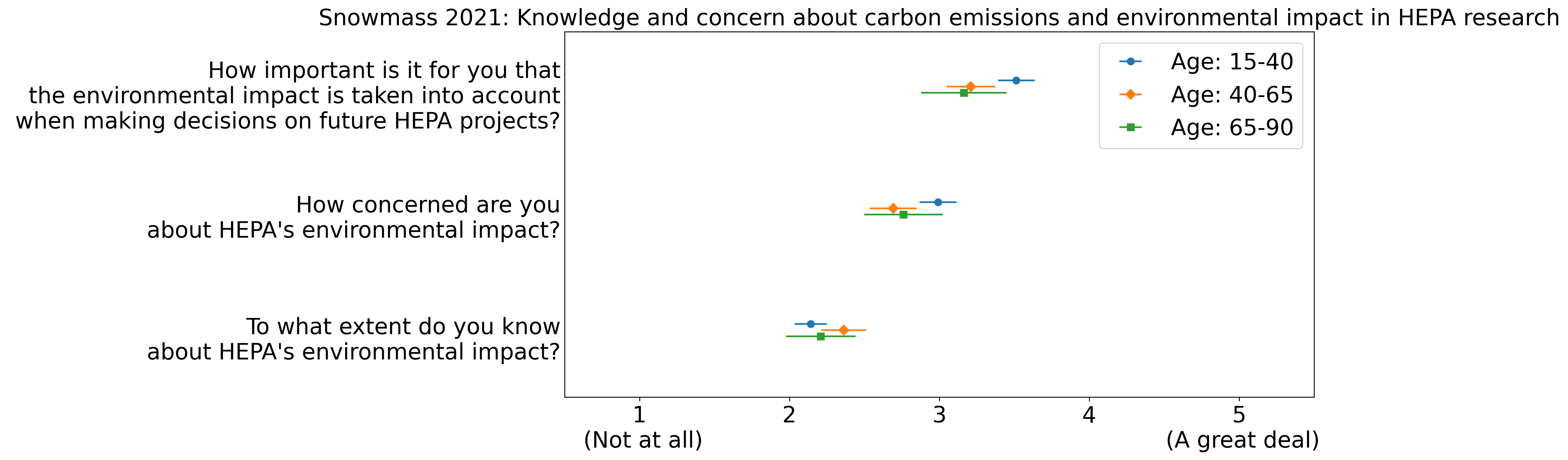}
  \caption{Perception of the community on the HEPA community’s environmental impact.}
  \label{fig:env}  
\end{figure}
\clearpage
\subsection{Survey Recommendations}
\label{sec:survey_rec}
The remainder of the survey focused on demographics, careers, workplace culture, diversity and racism, caregiving responsibilities, and the impacts of COVID-19. Here we discuss several recommendations based on the data presented in the Snowmass 2021 Community Survey Report. The survey team gathered, analyzed, and synthesized the data. Their synthesis included a discussion of the data that included some of the key findings on the current state of the field and potential areas for improvement. The recommendations in this section are derived from the survey team's discussion, and several members of the SEC survey team have helped develop this section.

In examining human experiences in the 2021 survey, the survey team found so much honesty and openness within the Snowmass community. People were honest about how much the COVID-19 pandemic has affected their career and personal life; they were honest about the experiences they faced involving harassment and racism; and they were honest about where they think the field is heading over the next decade. The survey uncovered unique perspectives and experiences for different career stages, primary workplaces, caregiving responsibilities, and across gender and racial lines. Yet the survey also revealed many similarities among the members of the Snowmass community, including concerns about salary, HEPA's environmental impact, and directions in which they would like the field to progress. Finally, many 2021 survey respondents reported feeling a lack of support in various places -- between peers and work colleagues, from an advisor, at the institutional level, or by the field as a whole. We each face a unique set of challenges for however long we work in HEPA, and on an individual level, we can extend more sympathy and understanding to our immediate working group, students, and other collaborators. However, individual cultural changes are not enough -- systemic changes need to be made by institutions and funding agencies to provide effective support for everyone, especially for early career scientists and those with caregiving responsibilities. By exploring the varied experiences of scientists in and out of HEPA, the 2021 survey exposed many concerns and disparities that should be addressed through meaningful action and cultural changes. At the time of the next Snowmass process, the next survey can be used to check for improvements in these areas.

\begin{figure}[h!]
    \includegraphics[scale=0.5]{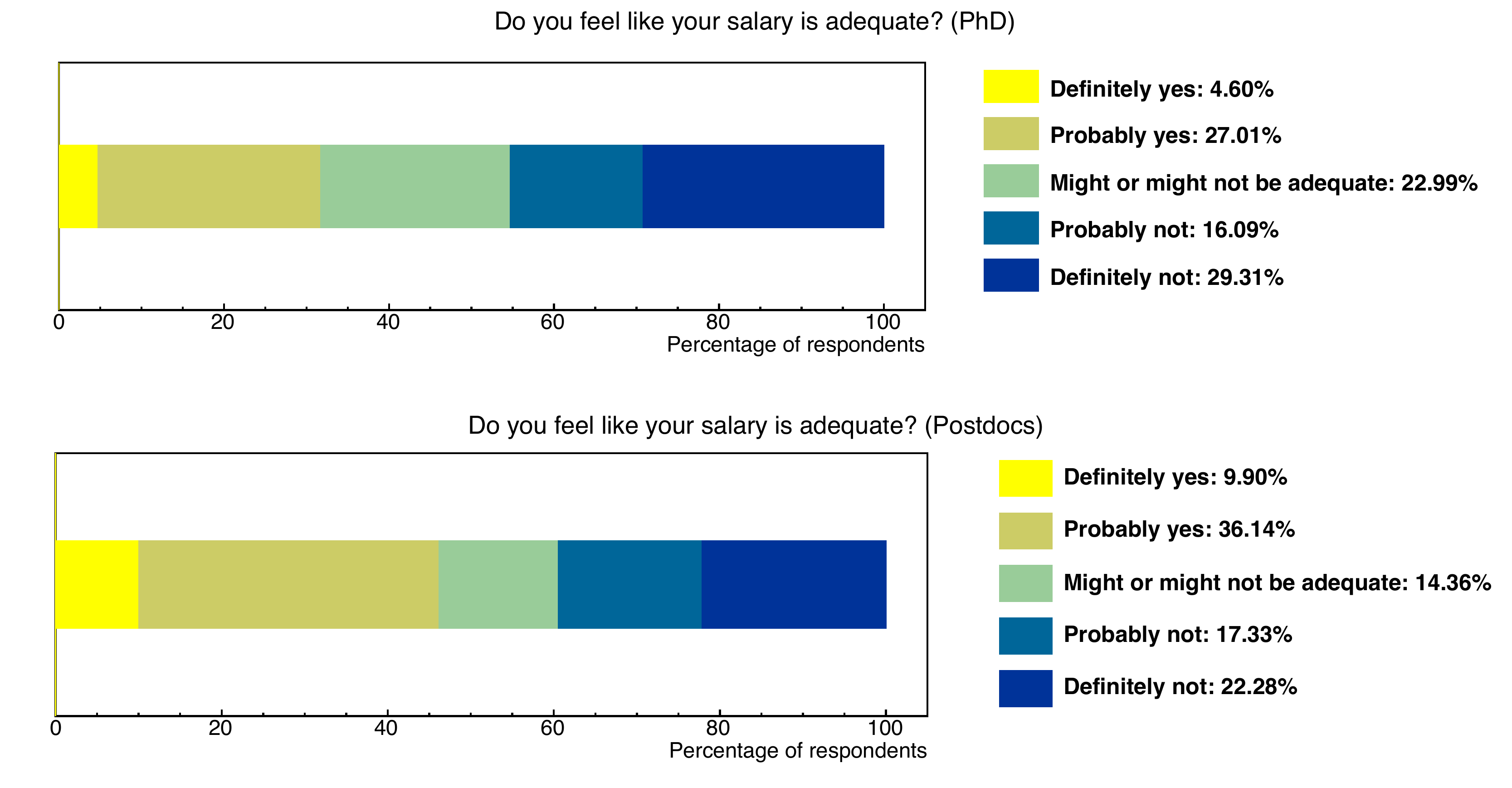}
    \caption{The survey team asked PhD students and Postdocs how they felt about their salary. The majority of PhD students (97.8\%) and Postdocs (99\%) answered this question.}
    \label{Q48_phd_postdocs_combined}
\end{figure}

\rule{\linewidth}{2.5pt}
\textbf{SEC Survey Recommendation 1: Institutions, funding agencies, and scientists should make systemic changes to provide early career scientists with adequate compensation.} \\
\rule{\linewidth}{2.5pt}
A large number of PhD students (45.4\%) and Postdocs (39.6\%) report feeling their salary probably or definitely is inadequate (Figure~\ref{Q48_phd_postdocs_combined}), indicating there might be a systemic issue with adequately compensating early career scientists for their contributions to HEPA. It is common for PhD students to be paid less for their contributions (e.g., due to the ``trainee'' or student aspect of the position), and there is little to no compromise for stipend amounts. Lack of compromise is also the case for Postdocs at some institutions, especially for institutions with high-demand positions or with strict salary requirements. Some Postdocs can negotiate their salaries, but those skills are not within the scope of the average U.S. graduate school curriculum.

Salary amounts should (at minimum) cover the cost of living for the specific institution's location, and they should increase every year to cover the cost of inflation. For universities, increasing stipends for graduate students might be handled outside of the physics department, so faculty and scientists can be advocates and consistently place pressure on the administrators who have control over determining graduate school stipends and Postdoc salaries. Funding agencies should also take cost of living and inflation into account while awarding fellowships to early career scientists and grants to faculty and scientists, so that research groups have adequate funds to compensate their students and Postdocs. Furthermore, graduate programs should offer training (e.g., in a class, seminar, or through a broader institutional effort) on how to negotiate salaries, and they should advertise or require the training at some point during the graduate curriculum. 

\rule{\linewidth}{2.5pt}
\textbf{SEC Survey Recommendation 2: Institutions and funding agencies should collect data on pay differences between junior and senior faculty and scientists and develop pay policies to address junior faculty and scientists' compensation concerns.}\\
\rule{\linewidth}{2.5pt}
When asked about their salaries, more tenure-track faculty think their salary probably or definitely is inadequate compared to tenured faculty, indicating a need to revisit and revise policies on faculty compensation. Similarly, when the survey team compared responses about salaries, scientists ranked the lowest (below senior scientists, tenure-track faculty, and tenured faculty) in terms of salary adequacy, and they also ranked higher than senior scientists in terms of salary inadequacy. The results continue to indicate that scientists in more junior positions are less satisfied with their salary, and the same sentiment is not shared with scientists in more senior positions. Conversely, more senior scientists reported that their salary is adequate at some level compared to the other three groups, and far fewer scientists and senior scientists reported that their salary is inadequate on some level compared to faculty. This difference could stem from the difference in pay between national labs and universities, and it indicates that national labs might be offering a more reasonable salary compared to universities.

\rule{\linewidth}{2.5pt}
\textbf{SEC Survey Recommendation 3: Institutions should track career outcomes and adequately train early career scientists to move into a variety of job sectors, especially industry positions.}\\
\rule{\linewidth}{2.5pt}
The 2021 survey results indicated less of a desire to remain in academic HEPA than to seek employment in another sector, both from the early career scientists currently in HEPA and the non-academics who recently departed HEPA for industry jobs. A majority of faculty and scientists also reported that they have thought about leaving HEPA or their institution. Even if scientists strive to remain in HEPA, they are predominantly concerned about job availability and already feel the competition surrounding them in their current workplace.  Institutions should provide early career scientists opportunities and training for professional development skills (e.g., negotiating salaries), so that they can effectively prepare for a variety of job sectors beyond HEPA. Funding agencies can also create opportunities for early career scientists to collaborate with industry partners, offering a bridge between HEPA and industry while also exposing early career scientists to a larger diversity of job sectors and mentors.

The survey shows that early career scientists with recent job offers are largely staying in HEPA. Overall, the vast majority of early career scientists who accepted or plan to accept a position applied for (and accepted) Postdoc positions. Even if PhD students ultimately accepted a non-academic position, they likely also applied for academic positions. While PhD students collectively applied for academic positions, Postdocs were more varied in their job search, where about two-thirds reported applying for academic and non-academic jobs alike. Compared to graduate students, more Postdocs indicated they were switching to a different job sector.

Early career scientists who are applying for jobs are largely interested in remaining in HEPA. Overall, early career scientists expressed much more interest in applying for academic positions versus industry positions. The results indicate that universities might not adequately advertise or train their students on how to apply for industry jobs, as the survey team found that early career scientists located outside of universities were more likely to apply for STEM industry positions. When asked to comment on why they chose their top sector, many early career scientists who rated academic positions highly wrote about their love for teaching and research. There was also a group of early career scientists who wrote about their struggles to remain in HEPA (e.g., not enough permanent positions, the expectation to complete multiple postdocs, etc.), and others who expressed dissatisfaction with HEPA and a desire to switch to a different sector. The early career scientists who took the 2021 survey are aware about the lack of permanent positions and highly competitive nature of HEPA. While the majority continue to apply for academic positions, some early career scientists are disappointed, dissatisfied, and moving out of HEPA. Furthermore, early career scientists who are applying for jobs are deeply concerned about job availability regardless of the sector. On the whole, early career scientists indicated much more concern about academic job availability compared to the industry job sector. 

Overall, early career scientists are applying for non-academic jobs (even if they ultimately remain in HEPA); there exists a subset of scientists who are dissatisfied with HEPA and wish to switch job sectors; and early career scientists expressed a deep concern for academic job availability. Outside survey data found that early career scientists (specifically Postdocs) are passionate about their research, but they also acknowledge the challenges of low pay, work-life balance, and a competitive job market \cite{afonja_salmon_quailey_lambert_2021}; similar results have also been observed in other surveys conducted during the pandemic \cite{woolston_2020}. 

The 2021 survey also received a balanced mix of respondents with a variety of age ranges, with a group who recently left HEPA and another group who left decades ago. This allowed us to compare their responses to look for changes in HEPA's job market and culture. The results indicated that younger and recently departed non-academic respondents seem less interested in an academic future: Significantly more non-academic respondents over 40 years old reported that they attempted to find a job in HEPA, and fewer respondents who left HEPA recently reported looking for an academic job. These results are concerning for early career scientists in HEPA who are planning to apply or currently applying for jobs, especially with the reports from early career scientists in HEPA about their dissatisfaction with HEPA and desires to switch sectors. Surveys conducted outside of the Snowmass process show that Postdocs are concerned about their career prospects and lack of support from their supervisor during the pandemic~\cite{nature_news_2020}, and these sentiments could, at worst, lead to a generational abandonment of HEPA as a whole.  

Institutions should train their early career scientists to explore a variety of job sectors comfortably and confidently. According to a report from the Organisation for Economic Co-operation and Development (OECD), institutions should track career outcomes from their PhD students so that the U.S. part of HEPA can better understand job opportunities on a local level along with what challenges new PhD holders face while transitioning into the next stage of their career. Institutions should also provide opportunities (e.g. seminars or internships) to Postdocs so they can gain professional development skills \cite{oecd_2021}. According to the survey results, networking is critical for those who are planning to leave HEPA, so both institutions and funding agencies should create networking opportunities.

\rule{\linewidth}{2.5pt}
\textbf{SEC Survey Recommendation 4: Institutions, national labs, and senior members of HEPA should support their employee's struggles with U.S. visas and immigration policies by advocating for updated policies, a streamlined application process, and inclusive hiring processes.}\\
\rule{\linewidth}{2.5pt}
The survey revealed that Postdocs are affected more by visa restrictions compared to graduate students, and some Postdocs applying for jobs reported that immigration issues were an important concern for them. Our results indicate that immigration issues disproportionately affect early career scientists in other racial groups, while White early career scientists remain largely unaffected. Current U.S. visa policies are largely inadequate to support Postdocs' transitions into non-academic job sectors \cite{geng_2022}. Immigration concerns should always be taken into account while training Postdocs how to navigate various job markets. 

The survey also shows that more tenured faculty take place in hiring compared to tenure-track faculty, where around half report that past or present U.S. policies on visas had some negative effect on their hiring ability. Additionally, scientists on the whole reported that past or present U.S. policies on visas had some negative effect on their hiring ability, indicating that past and present U.S. policies on visas have affected hiring at both universities and national labs in similar ways.

\rule{\linewidth}{2.5pt}
\textbf{SEC Survey Recommendation 5: Principal investigators and institutions should offer comprehensive support to their early career scientists.}\\
\rule{\linewidth}{2.5pt}
PhD students and postdocs rely on the support of their advisors or supervisors, not just their work colleagues. Principal investigators should offer support to their early career scientists, whether that be advocating for institutions to offer comprehensive mental-health services; exhibiting flexibility and patience toward their students and Postdocs; or advocating for increased salaries or benefits for their students and Postdocs. Additionally, institutions should provide resources and policies to support a healthy workplace culture and mental health, including providing insurances that fully cover off-campus therapy.

\rule{\linewidth}{2.5pt}
\textbf{SEC Survey Recommendation 6: Institutions and funding agencies should further investigate changing cultural trends in HEPA, including changing job expectations and increasing competition in the job market, and adjust policy accordingly.}\\
\rule{\linewidth}{2.5pt}
The average tenure-track faculty who took our survey noted more time as a Postdoc and more time looking for faculty positions compared to their averaged tenured faculty counterpart. This indicates that scientists searching for permanent positions in HEPA may spend longer amounts of time as an early career scientist. Furthermore, tenure-track faculty respondents set themselves apart from their tenured colleagues by forming two distinct groups: Those who spent 3-4 years as Postdocs, and others who spent 6-7 years as Postdocs. This seems to indicate that some tenure-track faculty completed two or more postdocs before securing their faculty position. Tenured faculty respondents roughly reported 3-5 years as a Postdoc. These results hint at changes within HEPA; expectations for faculty have changed over the past three decades in places such as collaborations and departments and the job market is more competitive now compared to the recent past. Scientists and senior scientists reported spending similar amounts of time as Postdocs, and although the difference is not significant, scientists did report spending a couple more months looking for a job compared to senior scientists. Similar to tenure-track faculty, the results from scientists might indicate the high competition currently being experienced in the HEPA job market.

Another hint to HEPA's changing culture is that the majority of teaching, tenure-track, and tenured faculty would prefer a reduced teaching load on some level, while the limited number of retired faculty respondents reported that they would not have preferred a reduced teaching load. Perhaps teaching expectations have changed over the past three decades, or perhaps other responsibilities have taken priority over teaching. 

\begin{figure}[h!]
    \includegraphics[scale=0.83]{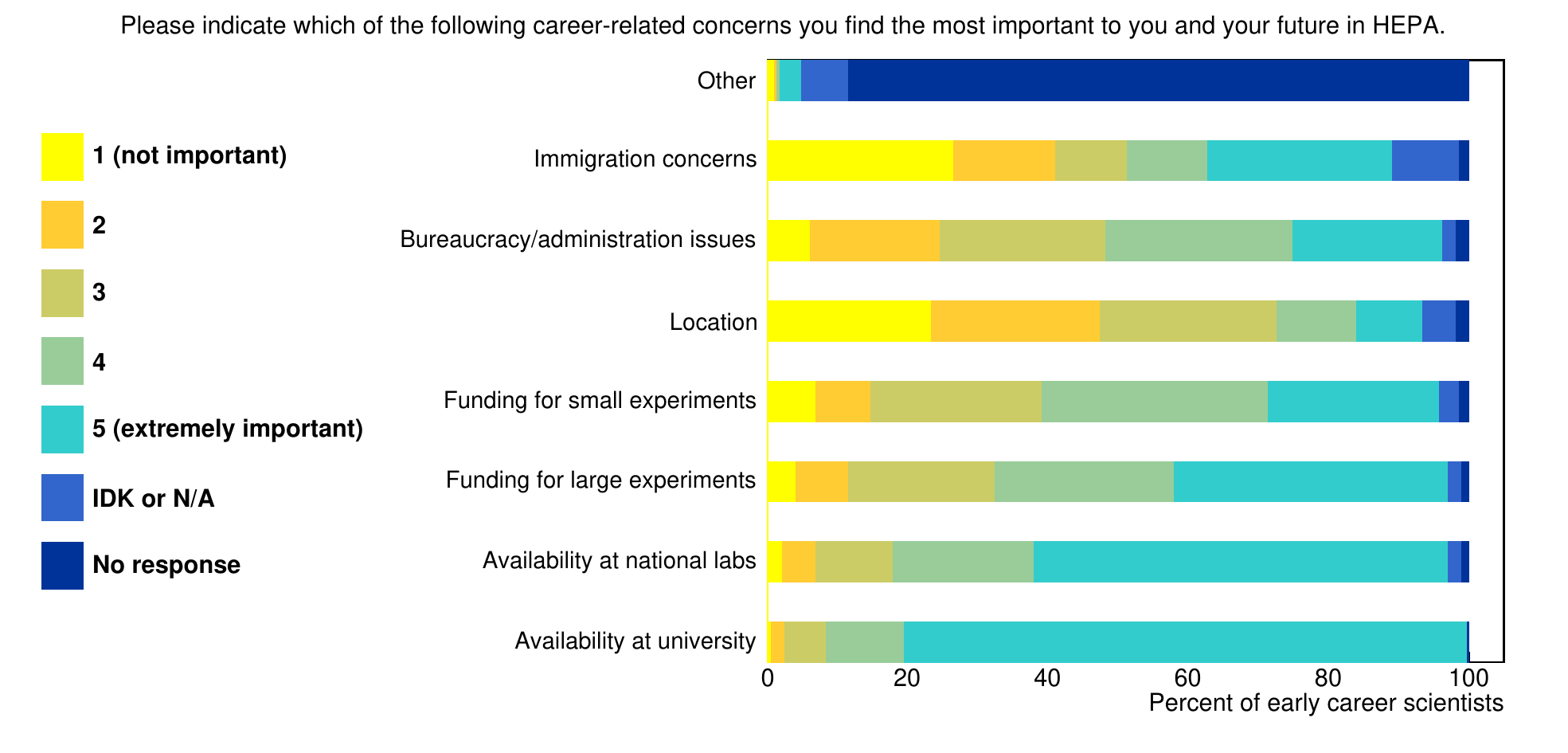}
    \caption{The survey team asked early career scientists who were likely to apply to academic positions about their career-related concerns.}
    \label{Q29_overall}
\end{figure}

\begin{figure}[h!]
    \includegraphics[scale=0.83]{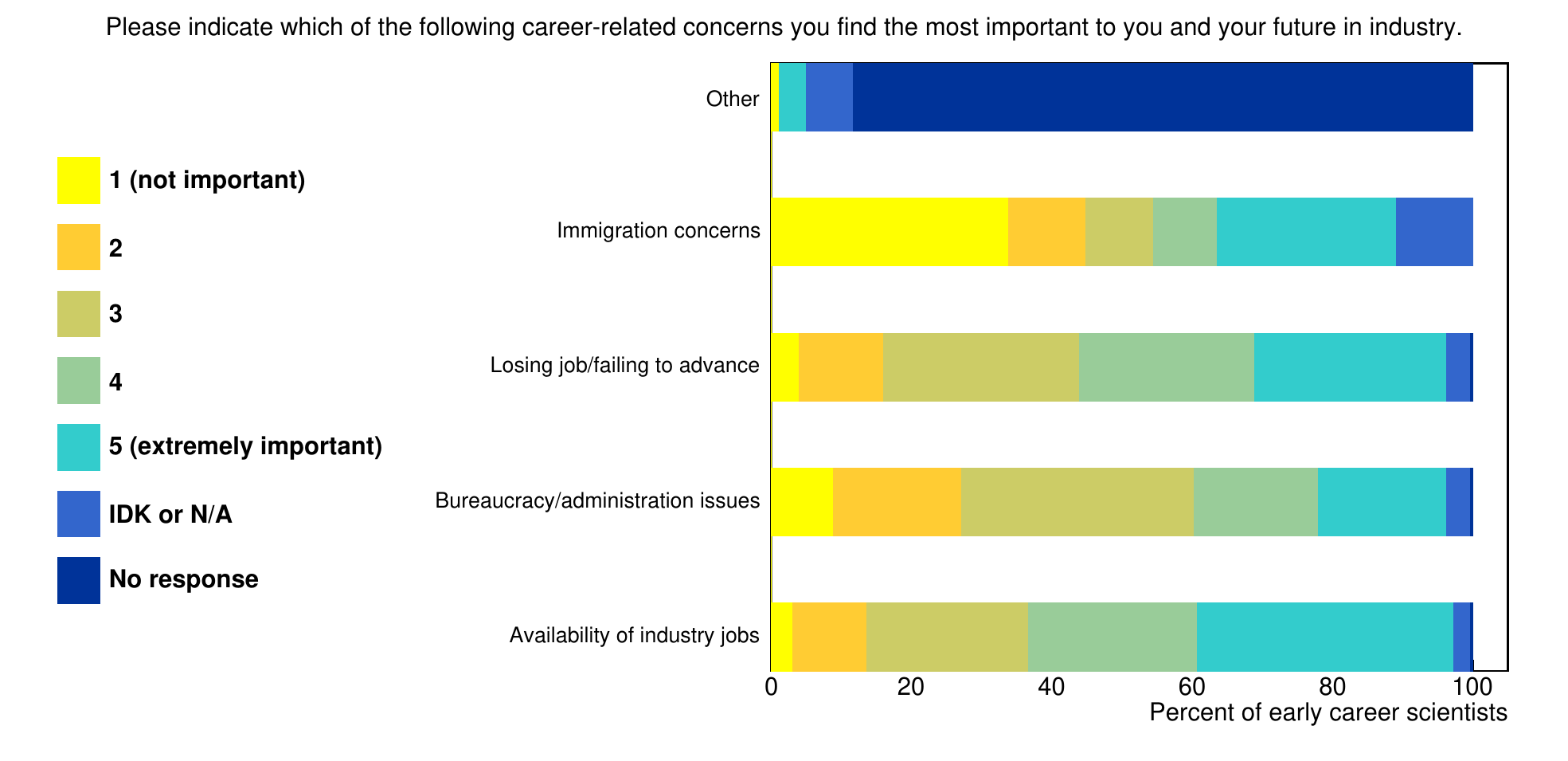}
    \caption{The survey team asked early career scientists who were likely to apply to industry positions about their career-related concerns. A little over half of the early career scientists surveyed saw this question.}
    \label{Q30_overall}
\end{figure}

Additionally, compared to faculty respondents, tenure-track faculty reported far fewer of their PhD students and Postdocs as employed after graduation. More tenure-track faculty also reported 0\% of their PhD students and Postdocs as employed. This could be related to early career scientists' concern over job availability, especially if both groups of scientists are applying for academic positions. Furthermore, the survey team discovered a similar report between scientists versus senior scientists for the PhD students and Postdocs they mentor. Because there is a difference between the responses of tenure-track and tenured faculty members, this once again indicates that HEPA's culture and job market have evolved over the past three decades. Figure~\ref{Q29_overall} further highlights the concern about the availability of positions at universities and national labs, especially when compared with industry positions (Figure~\ref{Q30_overall}).

While further investigation into these trends is needed to better quantify their sources and effects, it is clear that institutional and funding agency policies will need to change with the culture. For example, more long-term intermediate positions may be necessary given the extended timelines for career advancement, and assessment criteria may need updating to reflect changing expectations.

\rule{\linewidth}{2.5pt}
\textbf{SEC Survey Recommendation 7: Undergraduate students should be compensated for their research responsibilities.}\\
\rule{\linewidth}{2.5pt}
One concerning result from the survey was that about one-tenth of tenure-track faculty and one-fifth of tenured faculty reported not compensating their undergraduate students for their research responsibilities. This result is coupled with reports from some undergraduate student respondents that they are not compensated for their research responsibilities. Similar to unpaid internships, uncompensated research is an opportunity that is disproportionately unavailable to students from lower income backgrounds or those who lack financial support for their education expenses because they are unable to spend their time on uncompensated labor~\cite{rothschild_rothschild_2020}.

As one method to increase inclusiveness within HEPA, undergraduate students should be compensated for their research responsibilities, whether it is through financial compensation, course credit, outside fellowships, work study programs, or other means.  This may need to be handled on a department-by-department basis, but funding agencies should also allocate adequate funding toward compensating undergraduate students in faculty research groups. Funding agencies should also continue to fund fellowships and programs (e.g., the NSF REU program \cite{NSF_REU}) that support and nourish undergraduate student research opportunities.

\rule{\linewidth}{2.5pt}
\textbf{SEC Survey Recommendation 8: Institutions and funding agencies should provide support for outreach, mentoring, and advocacy efforts while continuing support for research, especially for underrepresented groups in physics. Institutions should include service work like mentorship, outreach, and advocacy in job expectations, ensure that faculty and scientists are given adequate time and credit for this work, and ensure that service work is equitably distributed.}\\
\rule{\linewidth}{2.5pt}
The survey results indicate that institutions are not systemically hindering research, outreach, or mentoring for their faculty members. On the other hand, faculty reported that their institutions help with research much more than outreach or mentoring. Furthermore, more faculty reported that their institution neither helps nor hinders mentoring and outreach compared to research. These results indicate that institutions are not actively helping with faculty outreach or mentoring efforts. Female faculty might also be disproportionately affected by their institutions compared to male faculty, as our results showed more female faculty reporting that their institution hinders their research on some level.

The vast majority of faculty work more than the standard 40-hour work week: A little less than half of the faculty respondents reported committing around 23 hours per week to research at the very least, and a quarter of faculty reported committing at least 23 hours a week to teaching. Overall, the average faculty respondent commits the most amount of time to research, fairly equal amounts of time committed to teaching and mentoring, and far less time to outreach and advocacy -- a hierarchy that seems reasonable but might be unbalanced in terms of actual time committed. When the survey team compared White faculty with faculty in other racial groups, they found that the latter group reported committing more time to research, outreach, and advocacy. Over a quarter of all faculty reported committing time to other career items -- mostly service and administrative work -- and female faculty reported spending more time on these responsibilities compared to male faculty. The faculty members commit similar amounts of time to their administrative duties and their teaching/mentoring, leading to an unbalanced set of commitments. 

When it comes to comparing time commitments, nearly two-fifths of scientists reported spending almost all of their time on at least one career item, one-fifth more than faculty who made the same report. With forty-three percent of scientists committing at least 22 hours per week to leadership and ninety percent of scientists reporting some leadership responsibilities, the results indicate that national labs provide ample opportunity for leadership opportunities. The survey team also found that scientists older than 50 years old commit less time to leadership compared to those younger than 50, a sensible and reasonable result. More scientists reported committing more time to leadership than research, but this isn't too concerning if these leadership responsibilities bear research-rich fruit. The survey team found differences between the amount of time scientists commit to career-related items across gender and racial lines -- particularly when it came to research, service, and leadership. Female scientists disproportionately reported spending more time on service work compared to male scientists; White scientists reported committing more time to leadership; and scientists in other racial groups reported committing more time to research.

With the other half of their 40-hour work week committed to research, perhaps it's not too surprising that scientists commit far less time to items outside of research and leadership. For example, scientists reported spending less time on mentoring compared to faculty reports, a result that is unsurprising, although dedicated funding for national lab-based student research opportunities could also help increase the mentoring opportunities for scientists. Scientists commit much less time to advocacy than all other career items over the course of a year. Scientists are employed as government contractors, so this may in part be due to less legal flexibility due to the Hatch Act of 1939.

Another hint of institutional differences arose when the survey asked about whether institutions help or hinder outreach and mentoring. Compared to faculty, more scientists reported that their institution hinders outreach and mentoring on some level. Nearly a quarter more female scientists reported that their institution hinders advocacy compared to reports from male scientists, and more male scientists indicated that they didn't know whether their institution helps or hinders advocacy efforts. This indicates that advocacy efforts might not be prioritized by national labs and might even be hindered in some cases, and this hindrance is disproportionately noticed by female scientists. 

\rule{\linewidth}{2.5pt}
\textbf{SEC Survey Recommendation 9: Funding agencies should move toward funding models that support increased collaboration like block grants to universities.}\\
\rule{\linewidth}{2.5pt}
Compared to other career stages, faculty reported the highest amount of competitiveness at their workplaces and the lowest satisfaction level with the support from their colleagues. While the majority of faculty reported feeling like their salary probably or definitely is adequate, about one-fifth of faculty feel the opposite. A significant portion of the faculty who responded to the 2021 survey face enormous time commitments and responsibilities at work and home while lacking support from their colleagues and the field overall. Funding agencies don't have to play a passive role in light of these results; for example, they could shift funds into block grants in which universities directly pay salaries to their faculty members \cite{woolston_2021}. This could reduce some of the competition faced by faculty who compete for grants, and it also reduces the workload and headache for faculty with administrative or grant/funding time commitments. 

\rule{\linewidth}{2.5pt}
\textbf{SEC Survey Recommendation 10: Institutions should improve how they support scientists with caregiving responsibilities, especially those most impacted by the effects of the COVID-19 pandemic. This includes encouraging reasonable work hours, providing adequate salaries, offering paid Medical and Family leave to all employees and supporting employees who use it, subsidizing or offering childcare to their employees, and fairly evaluating caregivers' drop in productivity in the context of current events in hiring and promotion committees.}\\
\rule{\linewidth}{2.5pt}
A significant portion of the survey respondents (including early career scientists) had some form of caregiving responsibilities (28.3\%). The majority of caregivers reported caring for children, although some respondents reported caring for seniors, a person with a disability or medical condition, or some combination of those three groups. Disproportionately more female caregivers reported they are sole or primary caregivers compared to male caregivers, while more male caregivers described their responsibilities as part-time caregiving. Female caregivers reported slightly less free hours per week on average and more female caregivers also indicated substantial career effects because of their caregiving responsibilities, including rearrangement of their work schedule, decrease of hours, or unpaid leave. These results seem to hint that female caregivers are not as widely supported within HEPA, and some of them face extra struggles like sole caregiving.

When asked about the level of support they receive in academia as a caregiver, the scientists who participated in this survey feel the most support from their peers or immediate working group, and they feel the least amount of support from funding agencies and the overall field. When broken down by primary workplace, results indicate that caregivers at universities and other institutions where caregivers deal more directly with funding agencies face an extra set of pressure and competition with no additional support to match.

While comparing caregivers' reports about their support from peers versus their place of employment, the survey team found that the responses formed two main groups: caregivers who reported feeling the same levels of support between their peers and their place of employment, and caregivers who reported less support from their place of employment. Recall that faculty also reported the lowest satisfaction level with the support from their colleagues when compared to other career stages -- a troubling result when it seems like some caregivers might need to rely on their peers at times when their place of employment does not provide adequate support. Additionally, the first group of caregivers contains those who reported feeling no support from either their peers or their place of employment, highlighting that institutions can continue to improve how they support scientists with caregiving responsibilities.  

The HEPA field overall is providing inadequate support to many of its caregivers, and according to the 2021 survey results, the lack of support disproportionately affects caregivers across different gender, racial, and age lines. Disproportionately more female caregivers reported feeling little to no support from their place of employment, funding agencies, and the field. More caregivers in other racial groups and more caregivers younger than 50 years old also reported feeling lack of support from funding agencies and the field. Keep in mind that the majority of all caregivers reported little to no support by the field; when broken down by gender, the same proportion of male and female caregivers respectively reported feeling some support as a caregiver by HEPA, and more male caregivers reported that they don't know whether they feel supported by the field. Clearly the field can extend more support to all its scientists with caregiving responsibilities. 

Caregivers in HEPA were profoundly impacted by the COVID-19 pandemic, and they continued to feel these impacts at the time they took the survey. The majority of all caregivers with children reported less overall productivity compared to pre-COVID levels; they also reported significantly more time and effort spent on caregiving responsibilities during the pandemic. More female caregivers, caregivers in other racial groups, and caregivers younger than 50 years old reported spending more time and effort on caregiving compared to pre-COVID time and effort. For caregivers of seniors or people with a disability or medical condition, half of these caregivers reported less overall productivity compared to pre-COVID levels, and over half reported more time and effort spent on caregiving during the COVID-19 pandemic. Caregivers also expressed similar sentiments when asked to comment about their experiences; they also expressed an inability to return to pre-COVID productivity levels despite their best attempts. Finally, caregivers did not widely report additional resources for childcare provided by their institutions during the COVID-19 pandemic.

When comparing caregivers with children in different age groups, the results revealed that more caregivers with younger children reported severe effects due to the COVID-19 pandemic, and caregivers with younger children (notably, young children in school) reported more time and effort spent on caregiving during the COVID-19 pandemic. The results revealed that more caregivers with older children reported the same overall productivity compared to pre-COVID levels; at the same time, more caregivers with children 14 years old or older -- or younger than 6 -- reported the same amount of time and effort spent on caregiving. Overall, more caregivers with children younger than 14 years old reported spending more time and effort on caregiving compared to pre-COVID time and effort. More female caregivers reported having preschool-aged children compared to male caregivers, while more male caregivers reported having children older than 5 years old. All of these results indicate that female caregivers in HEPA are disproportionately affected by the COVID-19 pandemic, impacting overall productivity levels and requiring more time and effort spent on their caregiving responsibilities. 

Several interesting differences arose when the survey team split caregivers' responses by different primary workplaces, notably universities and national labs. More caregivers at universities disproportionately indicated substantial career effects because of their caregiving responsibilities. When broken down by career stage, faculty caregivers reported more or substantial effects due to their caregiving responsibilities compared to scientist caregivers. These results indicate that universities might not be providing adequate support for scientists with caregiving responsibilities.

Faculty and scientists (which made up a large portion of the respondents with caregiving responsibilities) reported working long hours, with significant time commitments toward a variety of career-related items. The average caregiver reported being free of all caregiving responsibilities for about 34 hours per week (far fewer hours than what the majority of faculty and scientists spend working per week), and caregivers at national labs reported several more free hours compared to caregivers at other institutions. Around a quarter of faculty and one-fifth of scientists reported that their salaries are probably or definitely inadequate -- so there's room for institutions to encourage reasonable work hours and increase salary, two effective methods of supporting those with caregiving responsibilities \cite{collins_2021}. 

The average respondent with caregiving responsibilities has access to some form of Family and Medical leaves from their employer, and they are not not suffering any professional setbacks in the event they utilized Family and Medical leave. While the caregiver might have access to Family and Medical leaves, the leaves might be unpaid; some caregivers also reported that their institution does not offer Family and Medical leaves. More caregivers at universities and white caregivers reported that they don't know if their place of employment offers Family and Medical leaves. Disproportionately more male caregivers reported that they didn't know when it came to a few caregiving topics: More male caregivers didn't know if their place of employment offers Family and Medical leaves, whether their employer offered leaves during a time when they could have taken leaves, or whether they suffered any professional setbacks due to their taking of leaves. Caregivers also reported some professional setbacks while utilizing Family and Medical leaves, and this was disproportionately reported by female caregivers, caregivers in other racial groups, and caregivers at national labs. Institutions can offer paid Family and Medical leaves as one means of supporting the scientists with caregiving responsibilities \cite{collins_2021}, and institutions can actively encourage and support the scientists who need to utilize the leave.

With the COVID-19 pandemic predominantly affecting access to childcare and productivity levels, caregivers in HEPA require more support than ever from all angles: Work colleagues, department or institution administrators, and funding agencies. While specific action can be taken by the U.S. government concerning caregivers in all U.S. job sectors \cite{collins_2021}, institutions do not need to wait to implement such changes. Institutions should offer adequate salaries and paid Family and Medical leaves for all employees. Institutions can also support caregivers by subsidizing or offering childcare to their employees. Hiring and promotion committees should fairly evaluate caregivers' drop in productivity in the context of current events. Scientists with caregiving responsibilities make valuable contributions and bring an irreplaceable perspective to HEPA; they should not be left behind by the field. 

The responses and comments from caregivers were deeply honest, personal, and sometimes heartbreaking. We should not ignore or dismiss the real and valid responsibilities that scientists in HEPA face alongside their research commitments.

\rule{\linewidth}{2.5pt}
\textbf{SEC Survey Recommendation 11: Institutions and funding agencies should evaluate and assess the impacts of the COVID-19 pandemic over the coming years and adapt policy to support those most affected.}\\
\rule{\linewidth}{2.5pt}
The COVID-19 pandemic had an inseparable effect on all of the results of the 2021 survey. In addition to the direct effects on the Snowmass process (both in the timeline and the nature of the work), the pandemic fundamentally altered the every day lives and work of all the members of the HEPA field. Our results found that these effects---some temporary, others permanent and career-altering---had varying impacts across different groups and career stages, while the levels of support offered by institutions were similarly varied. Caregivers in particular were heavily impacted, and it may be a few years before the true extent of the impacts of the COVID-19 pandemic can be properly assessed.

\rule{\linewidth}{2.5pt}
\textbf{SEC Survey Recommendation 12:  Institutions should adapt workplace policy to be more flexible with work from home in the future.}\\
\rule{\linewidth}{2.5pt}
The COVID-19 pandemic changed working arrangements, and the survey inquired about current and preferred teleworking arrangements. This survey was conducted over several weeks and asked respondents to tell us their current teleworking arrangement. The survey team found that $69.2\%$ of respondents spent a majority of their time working from home. Going forward, a large group, $44.5\%$ prefer to work $1-2$ days from home, while $26.7\%$ prefer their workplace, and $9.25\%$ would prefer to work only from home. This could indicate a shift in how the workplace is utilized for members of the field in the future.



\clearpage


\section{SEC at the Community Summer Study}
\label{sec:css}

The Snowmass Community Summer Study (CSS) took place from July 16 to 26, 2022. 
It was held in hybrid format and hosted by the University of Washington, with in-person activities held on the Seattle campus. 
The CSS planning committee included two early career members who served as representatives of SEC and the community at large.

SEC organized a number of sessions at the CSS, with a workshop-wide plenary that contained talks on the SEC core initiatives, survey report, and long-term EC organizations across HEPA.
Early career physicists took the lead in many other areas, including several successful events focused on industry careers, networking, and perspectives.
They also hosted community discussions on mental health and invisible disabilities and were represented in panel discussions on community topics such as COVID-19 and career development.
The level of interest in the early career perspective for Snowmass and the future of SEC led to the scheduling of an additional feedback session on early career issues in the final days of the meeting.
Various informal social gatherings also took place to connect early career scientists beyond professional capacities.

Early career CSS participants were vocal and enthusiastic about the discussion of physics drivers and future experimental facilities. 
They were contributors to white papers across all frontiers, and gave talks in sessions. 
T-shirts advocating for a muon collider and pins for the Cool Copper Collider were introduced by early career scientists and seen on participants throughout the CSS. 

A form was issued to the early career community following the CSS, to provide a last opportunity to share feedback and impressions. 
The form requested the user's academic/career level and gave the option to select one of five areas of feedback: on the Community Summer Study content or discussion, on content or discussion that overlaps with a frontier report, on recommendations emerging from SEC survey data, on personal input or input from other colleagues, or perspectives on future physics drivers and facilities.
A field was also provided to indicate the target audience of the feedback, namely the field (all members), the field (senior members), funding agencies, early career organizations, or other.

A total of 33 form responses were recorded.
These primarily came from postdocs and on the topic of future physics perspectives, which called out support for next-generation cosmology experiments. 
Considerable discussion took place within the EC community on the role of voiced support for specific physics facilities, and the amount of care that should be used to disseminate these opinions. 
A numerical poll was not distributed to avoid the biasing of large experiments or frontiers over smaller ones.

\clearpage
\section{Recommendations}
\label{sec:recs}

The recommendations in this section were prepared by the HEPA EC community and are a supplement to the survey recommendations in Sec.~\ref{sec:survey_rec}. They include recommendations from contributed white papers and community feedback obtained throughout the Snowmass process. 

\subsection{Increasing Early Career Representation in Decision-Making Bodies}

\rule{\linewidth}{2.5pt} 

\textbf{SEC Recommendation 1.1 - Early career representatives should be included in HEPA community-wide decision-making bodies and advisory panels such as HEPAP and P5. }

\rule{\linewidth}{2.5pt} 

\textbf{SEC Recommendation 1.2 - Funding agencies should increase opportunities for early career participation in review panels.}

\rule{\linewidth}{2.5pt} 

\textbf{SEC Recommendation 1.3 -  HEPA experiments should pursue organizational change to include early career representatives in all decision-making bodies. }

\rule{\linewidth}{2.5pt} 

\textbf{SEC Recommendation 1.4 - The APS DPF Executive Committee should extend the term of the early career representative to match that of members at large and should increase the number of early career representatives in that body to two.}

\rule{\linewidth}{2.5pt} 

\textbf{SEC Recommendation 1.5 -  The users community organizations such as USLUA, SLUO and the Fermilab Users Executive Committee should explicitly codify the inclusion of early career representatives. }

\rule{\linewidth}{2.5pt} 

\textbf{SEC Recommendation 1.6 -  HEPA Experiments should foster the existence of early career organizations and provide them with resources to pursue new initiatives. }

\rule{\linewidth}{2.5pt} 

\textbf{SEC Recommendation 1.7 -  DOE National Labs should continue to support early career groups and their participation in community-building activities where they exist and encourage their creation at labs where they do not yet exist.} 

\rule{\linewidth}{2.5pt}

\subsection{Addressing Accessibility and Economic Equity}

\rule{\linewidth}{2.5pt} 

\textbf{SEC Recommendation 2.1 - Funding agencies, universities, and DOE National Labs should work together to sustainably increase the pay grade of graduate students, postdoctoral fellows and others in equivalent career stages according to their skills in order to reach competitiveness with industry salaries and improve their quality of life. }

\rule{\linewidth}{2.5pt} 

\textbf{SEC Recommendation 2.2 - APS and/or DPF should increase the funds available for travel assistance to early career representatives on decision-making bodies to ensure equal access. }

\rule{\linewidth}{2.5pt} 

\textbf{SEC Recommendation 2.3 - Funding agencies should consider language that encourages participation of students and postdocs in decision-making bodies and other service to the field as part of the mentorship and retention component of their grant proposal evaluations. }

\rule{\linewidth}{2.5pt} 

\textbf{SEC Recommendation 2.4 - Funding agencies should further explore funding opportunities that may be available exclusively to early career individuals as principal investigators.
}

\rule{\linewidth}{2.5pt} 

\subsection{Robust Equity, Diversity, and Inclusion}
Early career scientists are disproportionately affected by issues of discrimination and harassment. Additionally, the power dynamics of between senior members in the field and early career scientists are particularly strong, which can exacerbate these concerns. 
The following recommendations are in line with those in the Community Engagement Frontier report, but have been drafted to address issues that disproportionately affect early career individuals. \\

\rule{\linewidth}{2.5pt} 

\textbf{SEC Recommendation 3.1 - Funding agencies should significantly increase the ability to investigate and address instances of harassment and discrimination directly as well as implement real consequences for grantees when necessary and without relying on reports from institutions.}

\rule{\linewidth}{2.5pt} 

\textbf{SEC Recommendation 3.2 - The field must significantly increase programs of in-reach, training, and awareness of the Americans with Disabilities Act (ADA) and its consequences for workplaces and professional environments.} 

\rule{\linewidth}{2.5pt} 

\textbf{SEC Recommendation 3.3 - Universities and national laboratories must implement and impart specific training to educate early career scientists on the step-by-step process of addressing and reporting instances of harassment and discrimination, as well as their potential consequences.} 

\rule{\linewidth}{2.5pt} 

\textbf{SEC Recommendation 3.4 - Large experimental collaborations and projects in both experimental and theoretical physics must continue to develop policies to investigate and address instances of harassment and discrimination directly as well as implement real consequences for collaborators when necessary.}

\rule{\linewidth}{2.5pt} 

\textsc{}

\subsection{Empowering SEC for the next Snowmass Process}

\rule{\linewidth}{2.5pt} 

\textbf{SEC Recommendation 4.1 -APS should provide additional dedicated funding to DPF during Snowmass years in order to support accessibility and increased participation of SEC.}
\rule{\linewidth}{2.5pt}

\textbf{SEC Recommendation 4.2 - DPF should formalize connectivity with early career leadership throughout APS, such as in other divisions that participate in Snowmass like the Division of Astrophysics (DAP), or in the Fora on Graduate Student Affairs (FGSA) or Early Career Scientists (FECS).}

\rule{\linewidth}{2.5pt}



\bibliographystyle{unsrt}
\bibliography{SEC/bibliography.bib}


\end{document}